\documentclass[aps,11pt,showpacs,superscriptaddress,amsfonts,amssymb,amsmath]{revtex4}

\usepackage{epsfig}
\usepackage{makeidx}
\usepackage{graphicx}
\usepackage{epstopdf}

\begin{document}

\title{Gravitational radiation of a vibrating physical string as a model for the gravitational emission of an astrophysical plasma}

\author{R.A.\ Lewis \footnote{Email address: r3l@psu.edu}}
 \affiliation{Penn State University (ret.), Boalsburg, PA , USA}
\author{G.\ Modanese \footnote{Email address: giovanni.modanese@unibz.it}}
 \affiliation{Free University of Bolzano, Faculty of Science and Technology, Bolzano, Italy}

\date{April 4, 2014}

\begin{abstract}

The vibrating string is a source of gravitational waves which requires novel computational techniques, based on the explicit construction of a conserved and renormalized (in a classical sense) energy-momentum tensor. The renormalization is necessary to take into account the effect of external constraints, which affect the emission considerably. Vibrating media offer in general a testing ground for reconciling conflicts between General Relativity and other branches of physics; however, constraints are absent in sources like the Weber bar, for which the standard covariant formalism for elastic bodies can also be applied. Our solution method is based on the linearized Einstein equations, but relaxes other usual assumptions like far-field approximation, spherical or plane wave symmetry, TT gauge and source without internal interference. The string solution is then adapted to give the radiation field of a transversal Alfven wave in a rarefied plasma, where the tension is produced by an external static magnetic field. Like for the string, the field strength turns out to be independent from the frequency. We give a preliminary example of a numerical solution based on parameters referred to Alfven waves in the solar corona. Further astrophysical applications require an extension of the solution procedure to second order in the amplitude, and consideration of border effects. Future work will also address numerical and analytical near-field solutions.

\end{abstract}


\maketitle

\section{Introduction} 

The gravitational radiation emitted by astrophysical sources can sometimes be evaluated from exact solutions of the Einstein equations, which include solutions in the strong-field regime, solutions with retardation effects in the source and solutions describing both the near-field and the far-field region \cite{Weinberg,Stephani}.
The exact solutions, however, do not cover all possible cases of interest, and general perturbation methods have been developed. Such methods are similar to those employed for the Maxwell equations and are typically based on the multipolar expansion \cite{Maggiore}, involving several approximations and assumptions:

(a) Linearization of the field equations. 

(b) Far field approximation. 

(c) No retardation in the source. 

(d) Symmetry of plane wave or spherical wave. 

(e) Transverse-traceless (TT) gauge.

These approximations have of course a limited validity and in certain cases are not adequate. In our recent work \cite{Lew-Mod} we proposed a method which requires only the linearity assumption (a) and relaxes the others, and we used it to compute the emission of a Weber bar. In practice, we computed exactly the retarded potential integrals, working in the harmonic gauge, without imposing the TT gauge. We found an explicit expression for the metric tensor, and not only for the irradiated energy-momentum, and stressed the importance of appropriate boundary conditions imposed at the ends of the bar. We compared the response to the field  of a ``generalized'' two-rings detector with the response computed in the (local) TT gauge, and found that they are the same, because the longitudinal components of the metric which are present in the harmonic gauge have no physical effects on the detector. 

In this work we apply the general method of \cite{Lew-Mod} to a problem which is relatively new and physically interesting, namely the computation of the gravitational emission of a plasma wave. The wave is modelled through an analogy with an oscillating string, whose length is larger than the wavelength of the gravitational emission. We discuss the validity and the limitations of this analogy, considering the concrete example of an Alfven wave in the solar corona and we obtain analytical and numerical results for the far field and the near field (the latter to be published in a separate paper).

The crucial first step of our method is the explicit construction of a full conserved $T^{\mu \nu}$ tensor (Sect.\ \ref{s1}).
In the usual solution procedure of the linearized Einstein equations for far-field waves, the only relevant components of the source are the mass-energy density $T^{00}$ and its multipoles \cite{Weinberg,Maggiore}. This is true, however, only if all the approximations listed above are valid. If we are instead interested, for instance, into the near field or into sources with internal interference, we will need all the components of $T^{\mu \nu}$.

Furthermore, we would like to stress that even if one works in the multipole approximation and assigns the $T^{00}$ component in such a way that the conservation of total momentum and angular momentum is satisfied, the condition $\partial_\mu T^{\mu \nu}=0$ may require complex  internal forces between the system's components. Some of these  forces are the analogue of constraints in classical mechanics. In the context of General Relativity constraints may have unexpected consequences; in the case of the oscillating string, they give a major contribution to the radiation field.

The construction of our conserved stress tensor is described in Sect.\ \ref{s1}.  Since we cannot simply assume that the string is held in tension by a rigid support, large masses $M_l$, $M_r$ are placed at the ends of the vibrating string, to provide appropriate boundary conditions. A renormalization procedure involved in the limit for $M_l, M_r \to \infty$ is described in detail. In the end, the radiation field does not depend on $M_l$, $M_r$. Some terms in the stress tensor are based on expressions  appropriate to free particles;  the terms $T^{13}$ and $T^{33}$  are based on the usual description of elastic media, while $T^{33}$  and the boundary mass-energy density are determined for consistency.

Gravitational radiation generated by terms linear in the string wave amplitude is presented in Sect.\ \ref{far}.  Analytic integration of the retarded integral solution to the Einstein equation shows that the contributions of boundary conditions nearly cancel those of the interior of the string.  A short string, with length much smaller than a gravitational wavelength, can be treated as a current quadrupole. However, a naive appication of the quadrupole formula which does not take into account the string tension and the end-masses would lead to a wrong result and a grossly over-estimated radiation field.

Preliminary results from numerical integration indicate a component of the radiation field not included in analytical expressions of the first order in the oscillation amplitude.  (Parameters for the numerical integration presented in Sect.\ \ref{rad-pla} are based on properties of Alfven waves in the solar corona; a detailed description of the numerical integration method will be presented elsewhere.) However, attempts at constructing a conserved second-order $T^{\mu \nu}$ tensor appear to lead to inconsistencies, as discussed in Sect.\ \ref{Concl}. This might signal a fundamental conflict between the Einstein equations and systems which, like the oscillating string, feature ingredients alien to the foundations of General Relativity, as space-like correlations in stationary waves combined with external constraints.

Our method could also be applied in future to other sources for which the usual quadrupolar formula is not adequate. It could be used, for instance, to compute the radiation of an array of micro-bars oscillating at high frequency, devised for the laboratory generation of high-frequency gravitational waves (HFGWs). Such generators have been discussed, among others, by Baker and Woods \cite{Baker-Woods,Woods} and by Grishchuk \cite{Grishchuk1,Grishchuk2}. They are still far from concrete realization, but it has been estimated that they could generate HFGWs with amplitudes only 100 times smaller than the main astrophysical sources. Their high frequency implies that interference and retardation effects of the near field in the source would be important in that case.

\section{Construction of the conserved energy-momentum tensor}
\label{s1}

The definition of the energy-momentum tensor $T^{\mu \nu}$ for a complex elastic physical source can be based on a general formalism developed by Carter and Quintana \cite{Carter} and which has been applied to neutron stars \cite{Karlovini}. This technique, however, is not suitable for handling the emission of a vibrating string, which is characterized by the presence of a tension imposed by external constraints. (See also Sect.\ \ref{Concl}.) And while for electromagnetic waves only the motion of the charged parts of the source has importance and any other devices, interactions or constraints in the source, even if crucial for its motion, do not directly affect the emission, for gravitational waves any internal stress of the material which plays a role in determining the dynamics of the source may contribute significantly to the emission. 

\begin{figure}
\begin{center}
  \includegraphics[width=10cm,height=7cm]{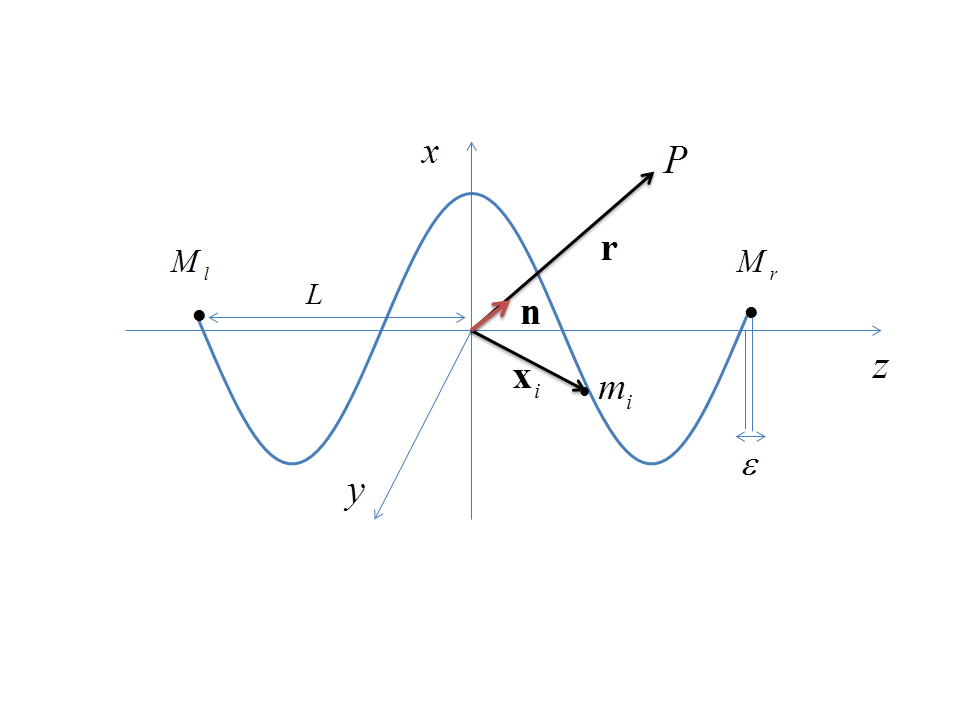}
\caption{Stationary wave on a string. The transverse oscillations take place in the $x$-direction. The string is held in tension by two large end-masses $M_l$ and $M_r$, placed very close to the first and last nodes of the oscillation. The distance $\varepsilon$ is an infinitesimal regulator, the masses are proportional to $\varepsilon^{-1}$. The radiation field is measured at the generic point $P$ with coordinate ${\bf r}$, and does not depend on $\varepsilon$ and on the end-masses in the limit $\varepsilon \to 0$. ${\bf n}$ is the unit vector of ${\bf r}$. In the integration procedure, the mass of the string is at first discretized into masses $m_i$ located at ${\bf x}_i$. } 
\end{center}
\label{f1}      
\end{figure}

The oscillation of our string can be described as a stationary wave with nodes on the $z$ axis (Fig.\ \ref{f1}) and displacement $u(z,t)$ in the $x$ direction given by
\begin{equation}
u\left( {z,t} \right) = {u_0}\cos \left( {\frac{{j\pi }}{{2\left( {L - \varepsilon } \right)}}z} \right)\cos \omega t, \qquad -L \leq z \leq L
\label{stat-wave}
\end{equation}
where $j$ is an odd integer related to the number of nodes and $\varepsilon$ is a regulator which eventually will tend to zero. ($\varepsilon$ has the dimensions of a length.) The presence of $\varepsilon$ in eq.\ (\ref{stat-wave}) means that the first and last nodes of the oscillation are not exactly at the extremes of the string, but approach them when $\varepsilon \to 0$.

In order to obtain a consistent energy-momentum tensor, we assume that the string is attached to two end-masses which hold it in tension. These ``left and right'' masses, denoted by $M_l$ and $M_r$, must be very large, because the string exerts a force on them and tends to pull them together. We expect that after a certain time this motion of the end-masses causes the string to relax; however, if the masses are very large, the relaxation time will be very long in comparison to the oscillation period. Actually, we suppose that the end-masses tend to infinity, and more precisely that they are proportional to $\varepsilon^{-1}$. The proportionality constant $\gamma$ must have dimensions mass$\times$length:
\begin{equation}
M_l=M_r=\gamma \varepsilon^{-1}
\end{equation}
We can suppose that $\gamma$ is a function of elementary constants, but the values of $M_l$, $M_r$ (and therefore the value of $\gamma$) will not appear in any final results. Like in a renormalization process, the physical quantities will be finite and independent from $\varepsilon$ and from $M_l$, $M_r$, in the limit $\varepsilon \to 0$. Terms of order $\varepsilon$, $\varepsilon^2$ ... will be disregarded, and so will be the static field of order $\varepsilon^{-1}$ generated by the end-masses, since we are only interested into the radiation field. The longitudinal $z$-coordinates of the masses $M_l$, $M_r$, are respectively defined by $z_l(t)=-L+w_{left}(t)$ and $z_r(t)=L+w_{right}(t)$. The functions $w_{left}(t)$ and $w_{right}(t)$ are infinitesimal of order $\varepsilon$. They satisfy certain consistency conditions, which will be discussed in Sect.\ \ref{nu-0-3}. Also the functions ${u_{left}}(t) = u( - L,t)$, ${u_{right}}(t) = u(L,t)$, which give the transversal $x$-coordinates of the two masses, are infinitesimal of order $\varepsilon$.

Summarizing, let us adopt for simplicity the following abbreviated {\bf notations}:
\begin{equation}
\begin{array}{l}
u = u(z,t)\\
{u_l} = {u_{left}}(t) = u( - L,t)\\
{u_r} = {u_{right}}(t) = u(L,t)\\
{w_l} = {w_{left}}(t)\\
{w_r} = {w_{right}}(t)
\end{array}
\label{conv}
\end{equation}
Note that $u_l$, $u_r$, $w_l$ and $w_r$ are all infinitesimal of order $\varepsilon$. For example, $u_r$ could be expanded as follows:
\begin{equation}
u_r \simeq -u_0 \frac{j\pi \varepsilon}{2L} \sin \left( \frac{j\pi }{2} \right) \cos \omega t + O(\varepsilon^2)
\end{equation}
It shall also be understood that the delta-function $\delta(x-u)$ in the $x$-direction is always accompanied by a delta-function $\delta(y)$ in the $y$-direction.

\subsection{Setting up the general structure of $T^{\mu \nu}$}
\label{setting}

Let us choose the following first tentative expression for $T$:
\begin{equation}
{T_{str}^{\mu \nu }}\left( {{\bf x},t} \right) = \left( {\begin{array}{*{20}{c}}
{\rho {c^2}}&{\rho c\frac{{\partial u}}{{\partial t}}}&0&{{T^{03}}}\\
{\rho c\frac{{\partial u}}{{\partial t}}}&{\rho {{\left( {\frac{{\partial u}}{{\partial t}}} \right)}^2}}&0&{{T^{13}}}\\
0&0&0&0\\
{{T^{30}}}&{{T^{31}}}&0&{{T^{33}}}
\end{array}} \right)
\end{equation}
where $u$ is the transversal displacement defined in (\ref{stat-wave}), (\ref{conv}) and $\rho$ is the mass density.
The elements of $T^{\mu \nu}$ with $\mu=2$ or $\nu=2$ are zero because of the choice of the reference system; $T^{00}$, $T^{01}$ and $T^{11}$ are suggested by analogy with known cases \cite{Lew-Mod,Low,Weinberg,Maggiore} and the elements $T^{03}$, $T^{31}$ and $T^{33}$ will be defined below, in such a way to respect the conservation conditions $\partial_\mu T^{\mu \nu}=0$.

The mass density in the string can be expressed as the sum of a ``S-term'' (String) and a ``B-term'' (Boundary),
\begin{equation}
\rho \left( {{\bf x},t} \right) = \rho_S \left( {{\bf x},t} \right) +\rho_B \left( {{\bf x},t} \right)
\label{rho}
\end{equation}
 where $\rho_S$ is straightforward:
\begin{equation}
\rho_S \left( {{\bf x},t} \right) = \sigma \delta (x-u) \theta \left[ (L+z-w_l)(L-z+w_r) \right]
\label{rho_s}
\end{equation}
and $\sigma$ is the mass for unit length of the string. (Remember the conventions (\ref{conv}) and that a $\delta(y)$ is always understood.) The density $ \rho_S$ satisfies the  continuity equation $\frac{{\partial \rho_S }}{{\partial t}} + \nabla  \cdot \left( {\rho_S {\bf{v}}} \right) = 0$. The boundary density $\rho_B$ will be determined by imposing the conservation conditions $\partial_\mu T^{\mu \nu}=0$.

The definition of the elements $T^{31}$ and $T^{33}$ is based on standard elasticity theory (see for instance \cite{Kittel}). By analogy with the case of a Weber bar we set    
\begin{equation}
{T^{31}}\left( {{\bf x},t} \right) =  - {T_{str}}\frac{{\partial u}}{{\partial z}}\delta (x-u) \theta \left[ (L+z-w_l)(L-z+w_r) \right]
\label{T31}
\end{equation}
and
\begin{equation}
{T^{33}}\left( {{\bf x},t} \right) =  - {T_{str}}\delta (x-u) \theta \left[ (L+z-w_l)(L-z+w_r) \right]
\end{equation}
where $T_{str}$ is the tension in the string. The definition of the element $T^{30}$, which is a pure boundary term, will be given in Sect.\ \ref{nu-0-3}.

Note that the functions $w_l$ and $w_r$ can be omitted when the quantities $\rho_S$, $T^{31}$ or $T^{33}$ are multiplied by quantities of order $\varepsilon$, because, for instance,
\begin{equation}
\delta(L+z-w_l)=\delta(L+z)-w_l \delta ' (L+z) + O(w_l^2)
\end{equation}
 
\subsection{Implementing the conservation equation with $\nu=1$}
\label{nu-1}

Let us now impose the $\nu=1$ component of the conservation equation $\partial_\mu T^{\mu \nu}=0$, i.e.
\begin{equation}
\frac{{\partial {T^{01}}}}{{c\partial t}} + \frac{{\partial {T^{11}}}}{{\partial x}} + \frac{{\partial {T^{21}}}}{{\partial y}} + \frac{{\partial {T^{31}}}}{{\partial z}} = 0
\end{equation}
We note that $T^{21}$ is zero due to the choice of the reference system, and we disregard for the moment $T^{11}$ because it is of second order in the oscillation amplitude $u$. For the vibrating string (unlike for a Weber bar)  it turns out that effects which are quadratic in  the oscillation amplitude are important.  The conservation condition of the stress tensor must therefore be satisfied separately for first and second order terms, and we shall discuss the second-order contributions to $T$ separately. For now we concentrate on requiring that
\begin{equation}
\frac{{\partial {T^{01}}}}{{c\partial t}} + \frac{{\partial {T^{31}}}}{{\partial z}} = O(u^2)
\label{nu1}
\end{equation}
Using (\ref{rho}), (\ref{rho_s}) the first derivative is expanded as
\begin{equation}
\begin{array}{l}
\frac{{\partial {T^{01}}}}{{\partial t}} = \frac{\partial }{{\partial t}}\left( {\rho_S \frac{{\partial u}}{{\partial t}}} \right) + \frac{\partial }{{\partial t}}\left( {{\rho _B}\frac{{\partial u}}{{\partial t}}} \right) = \rho_S \frac{{{\partial ^2}u}}{{\partial {t^2}}} + \frac{\partial }{{\partial t}}\left( {{\rho _B}\frac{{\partial u}}{{\partial t}}} \right) + O(u^2)= \\
 = \sigma \frac{{{\partial ^2}u}}{{\partial {t^2}}}\delta (x-u) \theta \left[ (L+z-w_l)(L-z+w_r) \right] + \frac{\partial }{{\partial t}}\left( {{\rho _B}\frac{{\partial u}}{{\partial t}}}
\right) +  O(u^2) 
\end{array}
\label{old14}
\end{equation}
Note that a term involving $\frac{{\partial \rho_S }}{{\partial t}}$  is omitted in the expansion;  this term is proportional to $u^2$ and will appear in the second-order discussion. 
The $z$-derivative in (\ref{nu1}) leads, after insertion of (\ref{T31}), to three terms, one of which (involving the $z$-derivative of the $\delta (x-u)$  factor) is of second order in $u$, while the other two give
\begin{equation}
\begin{array}{l}
\frac{{\partial {T^{31}}}}{{\partial z}} =  - {T_{str}}\frac{{{\partial ^2}u}}{{\partial {z^2}}}\delta (x-u)\theta \left[ (L+z-w_l)(L-z+w_r) \right] + \\
 \ \ \  \qquad - {T_{str}}\frac{{\partial u}}{{\partial z}}\delta (x-u) \left[ \delta (L+z-w_l) - \delta (L-z+w_r) \right]+  O(u^2)
\end{array}
\label{old15}
\end{equation}
Comparing (\ref{nu1}), (\ref{old14}) and (\ref{old15}) at the internal points of the string, i.e. disregarding the term with the delta-function of $z$ in (\ref{old15}) and the term with $\rho_B$ in (\ref{old14}), we obtain the familiar string equation
\begin{equation}
\sigma \frac{{{\partial ^2}u}}{{\partial {t^2}}} - {T_{str}}\frac{{{\partial ^2}u}}{{\partial {z^2}}} = 0
\end{equation}
This equation leads to a standard relation between frequency, wavelength and propagation velocity in the string.  Another consequence is the following expression for the string tension:
\begin{equation}
{T_{str}} = \sigma \frac{{{\omega ^2}}}{{{{\left( {\frac{{j\pi }}{{2L}}} \right)}^2}}} = 2\frac{{{M_{str}}{\omega ^2}L}}{{{j^2}{\pi ^2}}}
\label{mass-tension}
\end{equation}

The conservation equation for the boundary terms is
\begin{equation}
\frac{\partial }{{\partial t}}\left( {{\rho _B}\frac{{\partial u}}{{\partial t}}} \right) - {T_{str}}\frac{{\partial u}}{{\partial z}}\delta (x - u)\left[ {\delta (L + z - {w_l}) - \delta (L - z + {w_r})} \right] = O({u^2})
\label{cons-B}
\end{equation}
This leads us to define the boundary density as follows:
\begin{equation}
\rho_B \left( {{\bf x},t} \right) = M_l \delta (x-u_l) \delta(L+z-w_l) + 
M_r \delta (x-u_r) \delta(L-z+w_r)
\label{def-rho-B}
\end{equation}
We now check explicitly that this definition of $\rho_B$ satisfies (\ref{cons-B}) and is consistent with the ``equations of motion'' of $u_l$ and $u_r$. From (\ref{cons-B}), (\ref{def-rho-B}) we obtain
\begin{equation}
\begin{array}{l}
\frac{{{\partial ^2}u}}{{\partial {t^2}}}\left[ {{M_l}\delta (x - {u_l})\delta (L + z - {w_l}) + {M_r}\delta (x - {u_r})\delta (L - z + {w_r})} \right] + \\
 - {T_{str}}\frac{{\partial u}}{{\partial z}}\delta (x - u)\delta (L + z) + {T_{str}}\frac{{\partial u}}{{\partial z}}\delta (x - u)\delta (L - z) = O({u^2})
\end{array}
\label{eq-B-1}
\end{equation}
Note that in the second and third term we have omitted $w_l$ and $w_r$ in the arguments of the delta-function of $z$ and that we have discarded a term in which the derivative $\partial / \partial t$ acts on a delta-function, because all these terms are of first order in the regulator $\varepsilon$.

In the second and third term of (\ref{eq-B-1}) we make explicit the action of the delta-function of $z$:
\begin{equation}
\begin{array}{l}
\frac{{{\partial ^2}u}}{{\partial {t^2}}}\left[ {{M_l}\delta (x - {u_l})\delta (L + z - {w_l}) + {M_r}\delta (x - {u_r})\delta (L - z + {w_r})} \right] + \\
 - {T_{str}}{\left. {\frac{{\partial u}}{{\partial z}}} \right|_{z =  - L}}\delta (x - u)\delta (L + z) + {T_{str}}{\left. {\frac{{\partial u}}{{\partial z}}} \right|_{z = L}}\delta (x - u)\delta (L - z) = O({u^2})
\end{array}
\label{eq-B-2}
\end{equation}
Recall that using the definition (\ref{stat-wave}) of $u(z,t)$ one has
\begin{equation}
{\left. {\frac{{\partial u(z,t)}}{{\partial z}}} \right|_{z = L}} =  - {u_0}\frac{{j\pi }}{{2L}}\sin \left( {\frac{{j\pi }}{2}} \right)\cos \omega t + O\left( \varepsilon  \right)
\label{eq-B-3}
\end{equation}
and that the same derivative, evaluated at $z=-L$, has the opposite sign. Finally, remembering that apart from terms infinitesimal in $\varepsilon$ we can disregard $w_l$ and $w_r$ in the arguments of the delta-functions, we conclude that the conservation equation for the boundary terms is verified, provided
\begin{equation}
{M_l}\frac{{{d^2}{u_l}}}{{d{t^2}}} = {M_r}\frac{{{d^2}{u_r}}}{{d{t^2}}} = {T_{str}}{u_0}\frac{{j\pi }}{{2L}}\sin \left( {\frac{{j\pi }}{2}} \right)\cos \omega t
\label{eq-B-4}
\end{equation}
This just tells us that the end-masses oscillate in the transversal $x$-direction with the same phase, responding to the force exerted by the string at their instantaneous location. Eq.\ (\ref{eq-B-4}) also confirms that the functions $u_l$ and $u_r$ are of order $\varepsilon$ (by recalling that $M_l$ and $M_r$ are of order $\varepsilon^{-1}$ and integrating in $t$).

\subsection{Conservation equations for $\nu=0$, $\nu=3$}
\label{nu-0-3}

The implementation of the conservation equations for $\nu=0$, $\nu=3$ follows similar lines as we have seen for the case $\nu=1$. First we define the last missing component $T^{30}$ of the energy-momentum tensor as a pure boundary term as follows:
\begin{equation}
T^{30}=M_l c \frac{dw_l}{dt} \delta(x-u) \delta(L+z-w_l) +
M_r c \frac{dw_r}{dt} \delta(x-u) \delta(L-z+w_r) 
\label{T30}
\end{equation}
Using this definition and the definitions of $\rho$ and $T^{33}$ given in Sect.\ \ref{setting}, we find that the conservation equations are satisfied provided
\begin{equation}
M_l  \frac{d^2w_l}{dt^2}-T_{str}=0 \qquad \qquad M_r  \frac{d^2w_r}{dt^2}+T_{str}=0
\label{T30-a}
\end{equation}
This just means that, as expected, the left end-mass accelerates to the right (i.e., towards the center of the string) under the effect of the string tension, while the right end-mass accelerates to the left.

\section{Computation of the far field metric}
\label{far}

\subsection{Component $h^{00}$: String term $h^{00}_S$}
\label{h-S}

The metric is computed to first order in $u$ through the Lienard-Wiechert formula. To this end, the vibrating string is treated as a set of mass points. Let us first consider only points internal to the string; the contribution of the boundary, i.e.\ of the ``regulator'' end-masses $M_l$ and $M_r$ is computed in the next subsection.

The contribution to the trace-reversed metric component $h^{00}$ of a point mass $m_i$ located at ${\bf x}_i$ can be written as
\begin{equation}
\bar h_i^{00}({\bf{r}},t) = \frac{{4G{m_i}}}{{{c^2}|{\bf{r}} - {{\bf{x}}_i}({t_i}')| - c{{\bf{v}}_i}({t_i}') \cdot \left[ {{\bf{r}} - {{\bf{x}}_i}({t_i}')} \right]}}
\label{2star}
\end{equation}
where $t_i'$ is the ``retarded'' time at which the wave has been emitted, that reaches the observation point ${\bf r}$ at the observation time $t$. The time $t_i'$ satisfies the equation
\begin{equation}
{t_i}' = t - \frac{1}{c}|{\bf{r}} - {{\bf{x}}_i}({t_i}')|
\label{2star-1}
\end{equation}
The expression (\ref{2star}) is both valid for near and far fields. For the present calculation, only far field radiation is considered. Standard binomial expressions are used to express $\bar h_i^{00}$ as a power series in $1/|{\bf r}|$. We also discard terms which are of second order in the oscillation amplitude $u$. For instance, the relevant terms for the distance $|{\bf{r}} - {{\bf{x}}_i}|$ are the following:
\begin{equation}
|{\bf{r}} - {{\bf{x}}_i}| = \sqrt {{r^2} + {\bf{x}}_i^2 - 2r{\bf{n}} \cdot {{\bf{x}}_i}}  \simeq r - {\bf{n}} \cdot {{\bf{x}}_i} + \frac{{{\bf{x}}_i^2}}{{2r}}
\label{2star-2}
\end{equation}
where we have denoted with ${\bf n}$ the unit vector of ${\bf r}$ and with $r$ the length of ${\bf r}$, i.e.\ ${\bf r}=r{\bf n}$. In the scalar product of the velocity ${\bf v}_i$ by ${\bf{r}} - {\bf{x}}_i$, the relevant term is
\begin{equation}
\left( {{\bf{r}} - {{\bf{x}}_i}} \right) \cdot {{\bf{v}}_i} \simeq r{\bf{n}} \cdot {{\bf{v}}_i}
\label{2star-3}
\end{equation}
because ${\bf{x}}_i \cdot {{\bf{v}}_i}$ is of second order in $u$.

With this approximation we obtain the following contribution of the mass $m_i$ to the metric:
\begin{equation}
\bar h_i^{00}({\bf{r}},t) \simeq \frac{{4G{m_i}}}{{{c^2}r}} + \frac{{4G{m_i}{\bf{n}} \cdot {{\bf{v}}_i}({t_i}')}}{{{c^3}r}}
\label{3star}
\end{equation}
The first term, however, represents a static field, and will be ignored. The oscillation velocity of the mass $m_i$ is only in the $x$-direction, and is the time derivative of the displacement $u(z,t)$, computed at the coordinate $z_i$ where the mass $m_i$ is at time $t_i'$:
\begin{equation}
{v_{i,x}}({z_i},{t_i}') = \frac{\partial }{{\partial t}}u\left( {{z_i},t - \frac{1}{c}|{\bf{r}} - {{\bf{x}}_i}({t_i}')|} \right)
\label{3star-1}
\end{equation}
Taking into account eq.\ (\ref{stat-wave}), where for the internal points (unlike for the boundary) we can take immediately the limit $\varepsilon \to 0$, we obtain to the desired approximation
\begin{equation}
{v_{i,x}}({z_i},{t_i}') \simeq  - \omega {u_0}\cos \left( {\frac{{j\pi {z_i}}}{{2L}}} \right)\sin \left( {\omega t + \frac{\omega }{c}r + {k_z}{z_i}} \right)
\label{3star-2}
\end{equation}
where ${\bf k}=\frac{\omega }{c} {\bf n}$ is the wavenumber vector for the gravitational wave. Note that since the string can be long compared to a wavelength, the phase $k_z z_i$ cannot be neglected.

The total field is given by a sum over mass points, which is converted to an integral with the substitution $m_i \to \sigma dz=\frac{M_{str}}{2L} dz$. The integral to be evaluated is 
\begin{equation}
{I_{str}} = \int\limits_{ - L}^L {dz\cos \left( {\frac{{j\pi z}}{{2L}}} \right)\sin (\Phi  + {k_z}z) = } 2\sin \left( {\frac{{j\pi }}{2}} \right)\frac{{j\pi L\cos ({k_z}L)}}{{{{(j\pi )}^2} - 4{L^2}k_z^2}}\sin \Phi 
\label{3star-3}
\end{equation}
where, here and in the following
\begin{equation}
\Phi = \omega t - k r 
\label{3star-4}
\end{equation}
The resulting metric element will be denoted with the subscript ``S'', to distinguish it from the boundary contribution:
\begin{equation}
\bar h_S^{00}({\bf{r}},t) =  - \frac{{8G{M_{str}}}}{{{c^3}r}}\sin \left( {\frac{{j\pi }}{2}} \right){n_x}\omega {u_0}\frac{{j\pi \cos ({k_z}L)}}{{{{(j\pi )}^2} - 4{L^2}k_z^2}}\sin \Phi 
\label{3star-5}
\end{equation}

\subsection{Component $h^{00}$: Boundary term $h^{00}_B$}
\label{h-B}

Each of the end-masses $M_l$ and $M_r$ generates a field which is given by the analogue of eq.\ (\ref{2star}). In this case, however, the displacements $u_l(t)$ and $u_r(t)$ are infinitesimal of order $\varepsilon$, while $M_l$ and $M_r$ are infinite of order $\varepsilon^{-1}$. Therefore in the series expansions we shall keep only the finite terms resulting from the product of a term $\varepsilon$ with a term $\varepsilon^{-1}$. The infinite term of order $\varepsilon^{-1}$ is the static field of the end-masses, and will be discarded.

The finite term resulting from the analogue of (\ref{2star}) is
\begin{equation}
\bar h_B^{00}({\bf{r}},t) =
\frac{{4G\sin \Phi }}{{{c^3}r}}\left[ {{M_l}{\bf{n}} \cdot {{\bf{v}}_l}({t_l}') + {M_r}{\bf{n}} \cdot {{\bf{v}}_r}({t_r}')} \right]
\label{cross}
\end{equation}
Remarkably, we can eliminate from this expression any dependence on $M_l$ and $M_r$. To this end, let us write the condition that the center of mass of the system ``string + end-masses'' is at rest at $x=0$. The average of the $x$-displacement of the string is
\begin{equation}
{{\bar u}_{str}} = \frac{{{u_0}}}{{2L}}\int\limits_{ - L}^L {dz\cos \left( {\frac{{j\pi z}}{{2L}}} \right)\cos \omega t = \frac{{2{u_0}}}{{j\pi }}\sin \left( {\frac{{j\pi }}{2}} \right)\cos \omega t} 
\end{equation}
The two end-masses oscillate with the same phase. We also assume that they are equal: $M_l=M_r=M$. Hence the center of mass is at rest if $2Mu_l = -M_{str} \bar u_{str}$, which implies
\begin{equation}
{M_l}\frac{{d{u_l}(t)}}{{dt}} = {M_r}\frac{{d{u_r}(t)}}{{dt}} = \frac{{{u_0}}}{{j\pi }}{M_{str}}\omega \sin \left( {\frac{{j\pi }}{2}} \right)\sin \omega t
\end{equation}
For plugging this into eq.\ (\ref{cross}), the different retarded times for the two masses must be taken into account. We obtain
\begin{eqnarray}
\bar h_B^{00}({\bf{r}},t) & = & \frac{{4G{M_{str}}}}{{{c^3}r}}\frac{{{u_0}\omega n_x}}{{j\pi }}\sin \left( {\frac{{j\pi }}{2}} \right)\left[ {\sin (\Phi  + {k_z}L) + \sin (\Phi  - {k_z}L)} \right] \\
& = & \frac{{8G{M_{str}}}}{{{c^3}r}}\frac{{{u_0}\omega n_x}}{{j\pi }}\sin \left( {\frac{{j\pi }}{2}} \right)\sin \Phi \cos({k_z}L)
\label{cross-2}
\end{eqnarray}
Finally, note that in passing from (\ref{cross}) to (\ref{cross-2}) we did not consider the $z$-components of the velocities ${\bf v}_l$ and ${\bf v}_r$, given respectively by the functions $\frac{dw_l(t)}{dt}$ and $\frac{dw_r(t)}{dt}$, of first order in $\varepsilon$ (cfr.\ Sect. \ref{nu-0-3}). This is because in the limit $\varepsilon \to 0$ and $M_l,M_r \to \infty$ the motion of the masses in the $z$ direction is very slow and the frequency of the radiation emitted by this motion tends to zero. It can be shown, however, that the power emitted in this longitudinal motion is finite. This issue will be analyzed in a separate paper. If the string is used as a model for a plasma wave, one can argue that this power is ultimately supplied by the source of the static magnetic field.

\subsection{Full far-field metric tensor, to first order in $u$}
\label{full}

The procedure followed for the computation of the component $\bar h^{00}$ can be repeated for the other components, producing the full far-field metric tensor (or better its time-dependent part)
\begin{equation}
\left[ \bar h^{\mu \nu}({\bf{r}},t) \right]  =  -\frac{{4G{M_{str}}}{u_0}}{L{c^3}r} \sin \left( \frac{j \pi}{2} \right)
\left[ \frac{\cos({k_z}L) (2{k}L)^2}{(j \pi)^2-(2{k_z}L)^2} \right]
\left( {\begin{array}{*{20}{c}}
{n_x n_z^2 }&n_z^2&0&0\\
{n_z^2 }&0&0&n_z\\
0&0&0&0\\
0&n_z&0&-n_x
\end{array}} \right)\sin (\omega t - k r) 
\label{full}
\end{equation}
This expression contains the $\varepsilon$-finite contributions of the string and of the boundary, both approximated to first order in the displacement $u$ (like the $T^{\mu \nu}$ tensor from which they are derived). The radiation metric (\ref{full}) satisfies the Lorentz-harmonic gauge condition. Note that the field strength is independent from the frequency.

The common factor in square bracket in (\ref{full}) can be rewritten as follows, in terms of the wave propagation velocity $v$ on the string:
\begin{equation}
F =\left[ \frac{\cos({k_z}L) (2{k}L)^2}{(j \pi)^2-(2{k_z}L)^2} \right]
= \left[ \frac{\cos({k_z}L) v^2}{c^2-v^2 n_z^2} \right]
\label{full-F}
\end{equation}
This factor shows clear interference effects in the source when $v$ approaches the speed of light (Fig.\ 2).

\begin{figure}
\begin{center}
  \includegraphics[width=10cm,height=7cm]{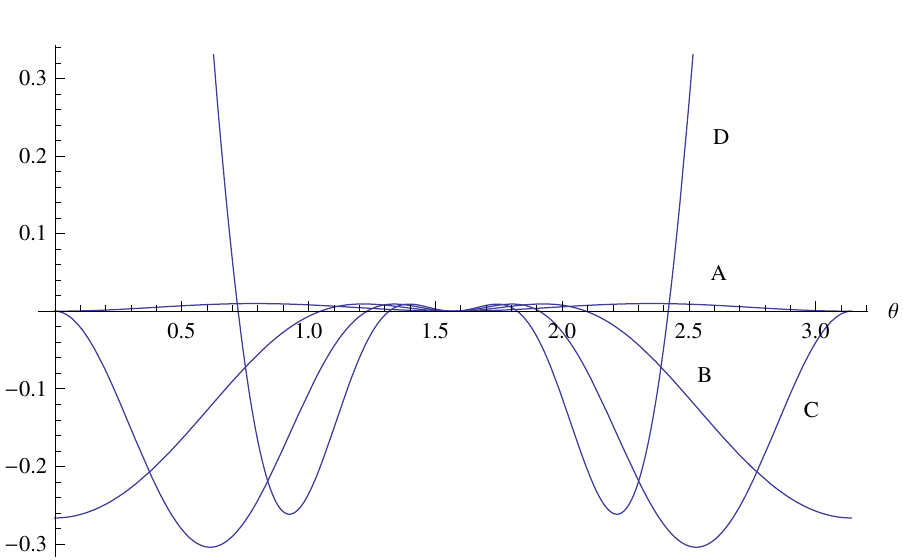}
\caption{Angular dependence of the radiation metric component $\bar h^{01}$, showing interference effects in the source (factor $n^2_z F$; compare eq.s (\ref{full}), (\ref{full-F})). The curves A, B, C, D refer to propagation velocities $v/c=0.2$, $v/c=0.4$, $v/c=0.6$, $v/c=0.8$, respectively. The node index is $j=5$. The effect of interference can be seen from the fact that minima and maxima in $\bar h^{01}(\theta)$ change their position in dependence on $v$. For different values of $j$, the shape of the low-velocity curve A is essentially unchanged, while the others vary markedly. The limit of $F$ for $\theta \to 0,\pi$ is finite and equal to $\frac{j\pi}{2}\sin \left( \frac{j\pi}{2} \right)$.} 
\end{center}
\label{interf}      
\end{figure}

\section{Radiation field of a transversal plasma wave}
\label{rad-pla}

As a source of gravitational waves, the oscillating string is clearly an interesting model, with features more general and complex than the Weber bar: it can have large transversal oscillations, it is a constrained system and a possible example of a source with length comparable with the emitted wavelength. The last feature produces interesting interference effects in the source (Fig.\ 2) but requires a medium with propagation velocity comparable with that of light, and this is not the case for any solid material. On the other hand, it is known that the velocity of transversal Alfven waves in a rarefied plasma can approach the speed of light. Alfven waves are a nice example of transversal oscillations of a material medium which do not require any mechanical constraint; the ``tension'' or restoring force is supplied by a background magnetic longitudinal field $B$. Their propagation velocity is given by $v=B/\sqrt{\mu_0 \rho}$, where $\rho$ is the mass density. A favorable environment for the propagation of fast Alfven waves is the solar corona. The waves are generated by solar flares at the border of the corona; their length can be of the order of $10^6$ m and their velocity up to $5 \cdot 10^7$ m/s \cite{McClements,Melrose}.

We now write the $T^{\mu \nu}$ tensor for an Alfven wave, to first order in the oscillation amplitude $u$. Some of its components ($T^{00}$, $T^{01}$ and $T^{13}$) are very similar to those of a material string, while others are different. The retarded integrals of the $T^{\mu \nu}$ components yield general first-order expressions for $\bar h^{\mu \nu}$ analogous to those of the string, and valid for a plasma wave of finite cross-section, in which boundary effects are disregarded (like for instance gradients of the transversal magnetic field at the border). 

With reasonable values for the physical parameters (see Table 1) we can compute the field numerically, not only in the radiation region, but also in the near field region, because in the numerical solutions it is possible to account more precisely for retardation effects. A detailed treatment of the numerical solutions will be given elsewhere. Also note that in the numerical computation of retarded integrals it is straightforward to take into account second-order components of the $T^{\mu \nu}$ tensor; but in view of the consistency problems with these components discussed in the Conclusions section, the numerical computations presented here are referred to the component $\bar h^{13}$, which has no second-order contributions.

\subsection{Construction of the $T^{\mu \nu}$ tensor for an Alfven plasma wave, to first order in $u$}

The wave is assumed to propagate in the $z$-direction and oscillate in the $x$-direction (Fig.\ 3). In the bulk of the wave the component $B_z$ of the magnetic field is constant up to second order in $u$. The component 
$B_y$ is zero and the component $B_x$ depends on $t$ and $z$, but not on $x$, like the plasma displacement $u(t,z)$. There are oscillating electric currents in the $y$-direction, because from the fourth Maxwell equation we obtain, disregarding the electric field, that $\partial_z B_x=\mu_0 J_y$.

\begin{figure}
\begin{center}
  \includegraphics[width=10cm,height=7cm]{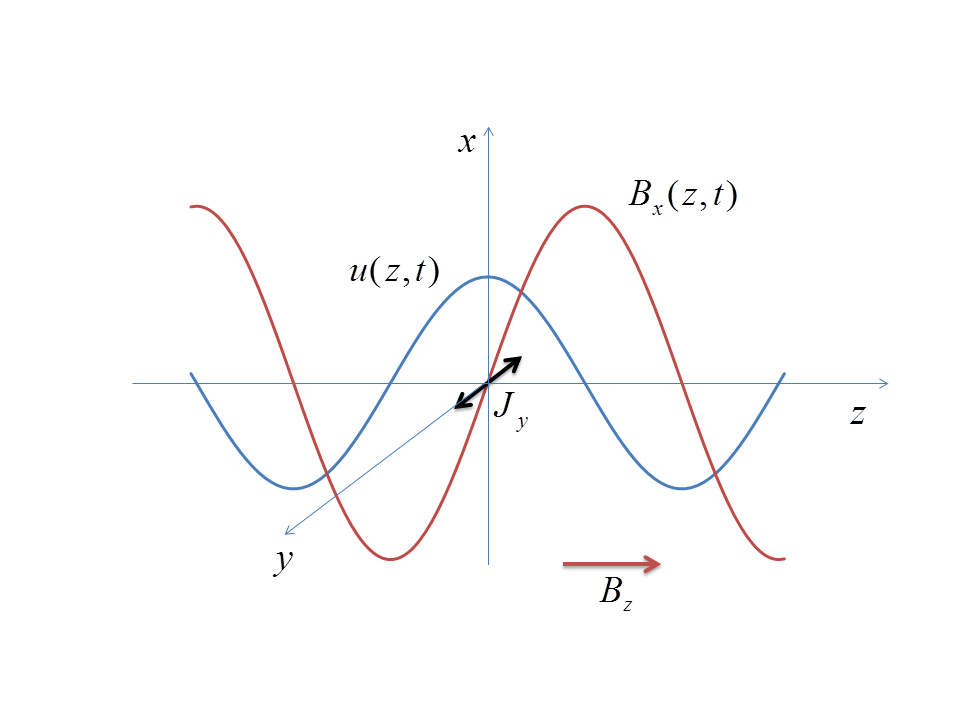}
\caption{Notation and coordinates for the description of a transversal plasma wave. $u(z,t)$ gives the collective displacement of the plasma in the direction $x$, $B_x(z,t)$ is the oscillating component of the magnetic field. $B_z$ is the static background field and $J_y$ shows the direction of the oscillating electric currents. } 
\end{center}
\label{plasma-wave}      
\end{figure}

The elements $T^{00}$ and $T^{01}$ are similar to those of the oscillating string, and contain the mass density $\rho$ and the time derivative of the displacement $\partial u/\partial t$. The element $T^{11}$ is of second order in $u$. The element $T^{03}$, which for the string is a pure boundary term containing the end-masses, is zero for the plasma to first order. The spatial components $T^{ij}$, which in the case of the string account for the elastic forces, are given for the plasma by the electromagnetic stress tensor. The only time-dependent element which does not vanish to first order is $T^{13}$, because $B_y=0$ and when $B^2_x$ and $B^2_z$ appear, one should remember that $B_x$ is of first order in $u$ (compare (\ref{pl-6})) and $B_z$ is constant in space and time, in the bulk of the wave. In conclusion, we can write
\begin{equation}
{T_{pl}^{\mu \nu }}\left( {{\bf{x}},t} \right) = \left( {\begin{array}{*{20}{c}}
{\rho {c^2}}&{\rho c\frac{{\partial u}}{{\partial t}}}&0&0\\
{\rho c\frac{{\partial u}}{{\partial t}}}&0&0&{\frac{{{B_x}{B_z}}}{{{\mu _0}}}}\\
0&0&0&0\\
0&{\frac{{{B_x}{B_z}}}{{{\mu _0}}}}&0&0
\end{array}} \right)
\label{pl1}
\end{equation}
The conservation equation $\partial_\mu T_{pl}^{\mu \nu }=0$ with $\nu=2$ is trivially verified, the one with $\nu=3$ gives $\partial_x (B_x B_z/\mu_0)=0$, which is also satisfied in our hypotheses. For $\nu=0$ we obtain a mass conservation equation. For $\nu=1$ we obtain the equation of motion of a small plasma volume under the  effect of the restoring force, according to the original Alfven equation
\begin{equation}
\rho \frac{{{\partial ^2}u}}{{\partial {t^2}}} =  - \frac{{{B_z}}}{{{\mu _0}}}\frac{{\partial {B_x}}}{{\partial z}}
\label{pl2}
\end{equation}
This equation can also be derived from the general equations of magneto-hydrodynamics, under suitable simplifying assumptions \cite{Jackson}.

Let us make the following ansatz for the instantaneous displacement $u(t,z)$ of the plasma and for the $x$-component of the magnetic field:
\begin{equation}
u(z,t)=u_0 \cos (k_j z) \cos \omega t
\label{pl3}
\end{equation}
\begin{equation}
B_x(z,t)=B_{0x} \sin (k_j z) \cos \omega t
\label{pl4}
\end{equation}
The wave vector $k_j$ is referred to the plasma wave with node index $j$, angular frequency $\omega$ and propagation velocity $v$:
\begin{equation}
{k_j} = \frac{{2\pi }}{{{\lambda _j}}} = \frac{{j\pi }}{{2L}} = \frac{{2\pi f}}{v} = \frac{\omega }{v}
\label{pl5}
\end{equation}
The functions (\ref{pl3}), (\ref{pl4}) satisfy eq.\ (\ref{pl2}), provided
\begin{equation}
\rho {u _0}{\omega ^2} = \frac{{{B_{0x}}{B_z}}}{{{\mu _0}}}\frac{{j\pi }}{{2L}}
\label{pl-6}
\end{equation}
Next consider a wave of finite extension in the directions $x$ and $y$, with cross-section $A$ (but still disregard the border effects). Multiply both sides of (\ref{pl-6}) by $A$ and replace $\rho A L$ by the total mass $M_{pl}$ of the oscillating plasma. This allows us to replace the quantity $B_{0x}B_z/\mu _0$ with an expression containing the mass of the plasma, the frequency $\omega$ of the wave, its oscillation amplitude $u_0$, its cross-section and the node index $j$:
\begin{equation}
 \frac{{{B_{0x}}{B_z}}}{{{\mu _0}}}=\frac{2Mu_0 \omega^2}{j\pi A}
\label{pl7}
\end{equation}
This shows that the field component $B_x$ is proportional to $u_0$, all the other parameters being fixed. From eq.s (\ref{mass-tension}), (\ref{pl7}) we can also obtain a relation between the magnetic field strengths and the tension of the string.

\subsection{Analytical formula for $\bar h^{13}$}

The retarded integrals of the components of the tensor $T_{pl}^{\mu \nu }$ are similar to those for the radiation field of the string, except for the necessity of a cut-off in the directions $x$ and $y$, to give the plasma wave a finite transversal cross-section. If this is done with a Gaussian function of the form $\exp \left( { - \frac{1}{2}{\chi ^2}k_x^2 - \frac{1}{2}{\zeta ^2}k_y^2} \right)$, the component ${\bar h}^{13}$ results, for example,
\begin{eqnarray}
{{\bar h}^{13}}({\bf{r}},t) & = & \frac{{ - 16\pi G}}{{r{c^4}}}\frac{{{B_{0x}}{B_z}}}{{{\mu _0}}}\frac{{{k_z}\chi \zeta }}{{k_j^2 - k_z^2}}\sin \left( {\frac{{j\pi }}{2}} \right)\cos ({k_z}L)\sin \Phi \exp \left( { - \frac{1}{2}{\chi ^2}k_x^2 - \frac{1}{2}{\zeta ^2}k_y^2} \right)  \nonumber \\
& =& \frac{{ - 16\pi G{M_{pl}}{u_0}}}{{r{c^5}L}}\frac{{{v^3}\cos \theta }}{{1 - \frac{{{v^2}}}{{{c^2}}}{{\cos }^2}\theta }}\sin \left( {\frac{{j\pi }}{2}} \right)\cos ({k_z}L)\sin \Phi \exp \left( { - \frac{1}{2}{\chi ^2}k_x^2 - \frac{1}{2}{\zeta ^2}k_y^2} \right)
\label{pl-analyt}
\end{eqnarray}
In the last expression we have replaced the product $\chi \zeta$ by the cross-section $A$, and eliminated $A$ and the magnetic field thanks to eq.\ (\ref{pl-6}). Remember that $\Phi=(\omega t - kr)$. Like for the string, the field amplitude is independent from the frequency and proportional to $M/L$.  A comparison between this formula and the result of a numerical evaluation of $\bar h^{13}$ is given in Fig.\ 4.

\begin{figure}
\begin{center}
  \includegraphics[width=10cm,height=5cm]{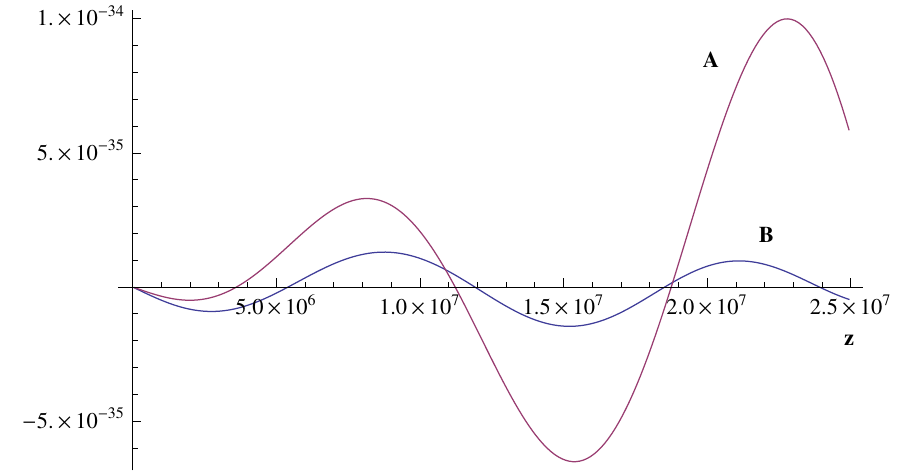}
\caption{{\bf A}: Numerical evaluation of the metric component $\bar h^{13}$ generated by a plasma wave along an arc at distance $r=2.5 \cdot 10^7$ m from the origin, as a function of the $z$-coordinate (with $y=0$). The parameters of the wave are given in Table 1. The phase of the time-dependent factor $\sin \Phi$ is equal to $5\pi/12$. The component $\bar h^{13}$ is the  retarded integral of $T^{13}$, which does not significantly depend on the square of the amplitude $u$. The result of the analytical computation (eq.\ (\ref{pl-analyt})) is not shown, because it is almost superposed to the numerical plot; there is a difference, however, which is shown amplified by a factor 10 ({\bf B}-curve). This difference can be attributed to a field component proportional to $1/r^2$, not present in the standard analytical $1/r$ far-field solution. Note that while for a Weber bar field components of this kind are negligible at a distance larger than one wavelength, in the present case they persist even at a distance of 80 wavelengths. } 
\end{center}
\label{h13}      
\end{figure}

\begin{table}
\begin{tabular}{ccc}
\hline\noalign{\smallskip}
 {\bf Parameter} & {\bf Symbol} & {\bf Value} \\
\noalign{\smallskip}\hline\noalign{\smallskip}
Mass of string, plasma & $M_{str}$, $M_{pl}$ & $5 \cdot 10^6$ kg \\
Half-length of string & $L$ &  $5 \cdot 10^5$ m \\
Oscillation mode & $j$ & 41 \\
Frequency & $f=\omega / 2\pi $ & 1000 Hz \\
Oscillation amplitude & $u_0$ & 22.1 m \\
String wavelength & $\lambda_j$ & 48780 m \\
Velocity of string waves & $v$ & $4.87 \cdot 10^7$ m/s \\
Grav.\ wave wavelength & $\lambda$ & $3 \cdot 10^5$ m \\
String tension & $T_{str}$ & $1.19 \cdot 10^{16}$ N \\
\noalign{\smallskip}\hline
\end{tabular}
\caption{Parameters for the numerical evaluation of the metric component $\bar h^{13}$ generated by a plasma wave, modelled as a string (Fig.\ 4).}
\label{table1}
\end{table}

\section{Conclusions and outlook}
\label{Concl}

This section collects the main open questions, to be addressed in future work. They include both physical and mathematical aspects, many of which have been already pointed out in the main text.

\begin{enumerate}

\item
{\it Correspondence of our general metric with the metric computed from the quadrupole current formula, in the limit of a string much shorter than the gravitational wavelength ($L \ll \lambda$).} This correspondence must take into account the mass renormalization caused by the constraint, which leads to a diminution of the effective mass by a factor of the order of $10^3$. With our choice of coordinates, the angular momentum density ${j^{jk}} = \frac{1}{c}\left( {{x^j}{T^{0k}} - {x^k}{T^{0j}}} \right)$ has non-zero ${j^{13}}$ components for the vibrating string. Gravitational radiation can be expressed in terms of the first moment of the angular momentum density. The series expansion of the factor $\cos(k_z L)$ in our general expressions gives terms representing current quadrupole, hexadecupole, etc. When $L$ approaches $\lambda$, a large number of multipoles is required, since the series of $cos(k_z L)$ converges slowly. 

\item
{\it Low-frequency emission from the constraints.} As mentioned in Sect.\ \ref{h-B}, in the limit when the constraining end-masses are very large, the frequency of their emission tends to zero, but the emitted power stays finite. Physically, this should be regarded as a real dissipation. Since actual constraints can supply a large but finite energy, this dissipation may be important in the total energy balance of the process.

\item
{\it Computation of numerical solutions valid at any distance, and correspondence with analytical solutions.} The numerical evaluation of the retarded integrals allows to take into account terms depending on all powers of $r^{-1}$, as we saw in the example of the ``far-near field'' component of $\bar h^{13}$ which persists even at distances much larger than one wavelength (Fig.\ 4). Another advantage of the numerical solutions is that they can include in a straightforward way the contributions from second-order terms in $T^{\mu \nu}$, after they have been identified unequivocally (compare Point 5 below). These contributions proportional to $u^2$ can also be identified through their time dependence, containing terms with frequency equal to twice the fundamental frequency. 

\item
{\it Analytical expressions for the near field.} These require a Fourier-Bessel solution to the Einstein equations. Analytic formulae for $\bar h^{00}$ and $\bar h^{01}$ near fields agree with numerical calculations, out to a radius of 20,000 meters.  The series expansions for the $K_0$ and $K_1$  Bessel functions converge slowly, therefore deviations between analytic and numerical results are attributable to truncation errors. 

\item
{\it Extension of the energy-momentum tensor to second order in $u$.}
Our explicit construction of the renormalized $T^{\mu \nu}$ tensor of the string is consistent with the conservation of $T^{\mu \nu}$ to first order in the oscillation amplitude. It is not yet clear, however, how to generalize this procedure univocally to second order. This is a challenging issue, as it often happens in classical and quantum field theory when passing from a linearized theory to one which is reliable also at higher orders. The feature of the string which most conflicts with General Relativity is probably the presence of the external constraints, whose dynamics is not included in the equations of the string itself. For comparison, in the case of the Weber bar the terms of order $u^2$ in $T^{\mu \nu}$ do not significantly affect the metric and are generally much smaller. The Weber bar can in principle be described through the general-relativistic formalism for elastic bodies of Ref.s \cite{Carter,Karlovini}. This formalism is equivalent to the introduction of an action of the form $S=-\int d^4x \sqrt{g} \rho$, where $\rho=n \varepsilon$ is the energy density, $n$ the particle density and $\varepsilon$ the energy per particle. A direct application of this formalism to the string appears to be impossible; compare also the discussion of energy flow in a string in \cite{Butikov} and ref.s.  A possible escape from the conflict is the introduction in General Relativity of additional degrees of freedom, like scalar fields or Chern-Simons fields, which allow to generalize the Einstein equations and the conservation equation of $T^{\mu \nu}$. Extensions of this kind have also been proposed as a basis for improved energy-momentum pseudo-tensors \cite{Marquet}.

\end{enumerate}

\end{document}